\documentstyle[aps,preprint]{revtex}
\def\mjpar{\mbox{${m_j^\parallel}$}}
\def\mjper{\mbox{${m_j^\perp}$}}
\begin{document}
\tightenlines
\draft
\title{Quantum transport in the cylindrical nanosize silicon-based MOSFET}
\author{S. N. Balaban$^a$, E. P. Pokatilov$^a$, V. M. Fomin$^{a,b}$, V. N. Gladilin$%
^{a,b}$, J.~T.~Devreese$^b$, W.~Magnus$^c$, W.~Schoenmaker$^c$, M. Van Rossum%
$^c$, and B. Sor\'ee$^c$}
\address{${}^a$Departamentul de  Fizica Teoretic\u{a},
Universitatea de Stat din Moldova,\\
MD-2009 Chi\c{s}in\u{a}u, Republica Moldova}
\address{${}^b$Theoretische Fysica van de Vaste Stof,
Universiteit Antwerpen (U.I.A.),\\
B-2610 Antwerpen, Belgium}
\address{${}^c$IMEC, B-3001 Leuven, Belgium}
\date{\today}
\maketitle
\begin{abstract}
A model is developed for a detailed investigation of the current flowing
through a cylindrical nanosize MOSFET with a close gate electrode.
The quantum mechanical features of the lateral charge transport are
described by Wigner distribution function which is explicitly dealing
with electron scattering due to acoustic phonons and acceptor impurities.
A numerical simulation is carried out to obtain a set of $I$-$V$
characteristics for various channel lengths. It is demonstrated that
inclusion of the collision term in the numerical simulation is important
for low values of the source-drain voltage. The calculations have further
shown that the scattering leads to an increase of the electron density in
the channel thereby smoothing out the threshold kink in $I$-$V$
characteristics. An analysis of the electron phase-space distribution
shows that scattering does not prevent electrons from flowing through
the channel as a narrow stream, and that features of both ballistic and
diffusive transport may be observed simultaneously.
\end{abstract}

\section{Introduction}

During the last decade significant progress has been achieved in the
scaling of the metal-oxide-semiconductor field-effect transistor (MOSFET)
down to semiconductor devices with nanometer sizes.
In Ref.~\onlinecite{Nakajima}, the fabrication of silicon quantum wires with
lengths and widths of about 60 nm and $\sim$~20~nm respectively is reported.
The conductance of those quantum wires was measured for a wide range of
temperatures, from 25~to~160~K.
The fabrication and the investigation of a 40 nm gate length n-MOSFET are
reported in Ref.~\cite{Ono}. The resulting nanosize n-MOSFET operates rather
normally at room temperature. Using nanoimprint lithography, a field effect
transistor (FET) with a 100~nm wire channel was fabricated \cite{Guo1} and
the characteristics of this FET at room temperature were investigated.
However, the small size of nanoscale MOSFET with a wide Si substrate
negatively influences the device characteristics due to the floating body
effect. Short channel effects together with random effects in the silicon
substrate are very well known to cause a degradation of the threshold
voltage and the appearance of uncontrollable charge and current in regions
far from the gate electrode. Therefore in a nanoscale conventional MOSFET,
the controlling ability of the gate electrode is substantially weakened.
Recently, considerable attention has been paid to SOI (Si-on-insulator)
MOSFETs, which are prospective for creating new nanosize devices.
In Refs.~\onlinecite{Woo,Joac1,Pid}, a MOSFET with very thin SOI was
theoretically investigated on the basis of a 2D analytical model, while in
Ref.~\onlinecite{Koh}, a 1D~model was used.
Drain-induced barrier lowering was considered \cite{Woo} and the physical
mechanisms which determine the subthreshold slope (S-factor) were analyzed
\cite{Joac1}. As a result, a substantial reduction of the short-channel
effect in the SOI MOSFET as compared to that in the bulk devices was
established. As was shown theoretically in Ref.~\onlinecite{Joac2}, the
use of a lightly doped source and drain leads to an increase of the effective
channel length what allows one to weaken the drain-induced barrier lowering.
In an SOI MOSFET with a Si-Ge source \cite {Ari}, an improved
drain-to-source breakdown voltage is achieved due to the absorption of
excess holes in the channel region.
In a transistor device with a channel sandwiched between oxide layers
(dual-gate MOSFET), the floating body effects are significantly suppressed
\cite{Sek}. A theoretical model of a dual-gate device is described in
Ref.~\onlinecite{Pik}. In Ref.~\onlinecite{Pok} we have investigated the
thermal equilibrium state of a nanoscale cylindrical silicon-based MOSFET
device with a close gate electrode (MOSFETCGE).
An advantage of the latter is the complete suppression of the floating
body effect caused by external influences. Moreover, the short-channel
effect in these devices can be even weaker than that in a dual-gate
structure.

The main goal of the present work is the investigation of quantum
transport in a nanosize MOSFETCGE device. We have developed a flexible 2D
model which optimally combines analytical and numerical methods and
describes the main features of the MOSFETCGE device. The theoretical
modeling of the quantum transport features involves the use of the
Wigner distribution function formalism~\cite{Frensley,Hideaki,Jensen}.
The paper is organized as follows. In Section~II, a description of the
system is presented in terms of a one-electron Hamiltonian.
In Section~III the quantum Liouville equation satisfied by the electron
density matrix is transformed into a set of one dimensional equations
for partial Wigner distribution functions. A one-dimensional collision
term is derived in Section~IV.
In Section V we describe a numerical model to solve the equations which have
been derived for the partial Wigner distribution function. In Section~VI
the results of the numerical calculations are discussed.
Finally, in Section~VII we give	a summary of our results and conclusions
about the influence of the scattering processes in nanosize MOSFETs.

\section{The Hamiltonian of the system}

We consider a cylindrical nanosize MOSFETCGE structure (Fig.~1) described
by cylindrical coordinates $(r, \phi, z)$, where the $z$-axis is chosen to
be the symmetry axis. In the semiconductor pillar the electron motion is
determined by the following Hamiltonian
\begin{equation}  \label{Hame}
{\hat H}_j=-\frac{\hbar^2}{2\mjper} \frac{\partial^2}{\partial^2 {\bf r}%
_\perp} -\frac{\hbar^2}{2\mjpar} \frac{\partial^2}{\partial^2 z}+V(%
{\bf r}),
\end{equation}
where $V({\bf r})=V_b({\bf r})+V_e({\bf r})$ is the potential energy
associated with the energy barrier and the electrostatic field,
respectively; $\mjper$ and $\mjpar$ are
the effective masses of the transverse (in ($x,y$)-plane) and
longitudinal (along $z$-axis) motion of an
electron of the $j$-th valley.

The electrostatic potential energy $V_e({\bf r})$ satisfies Poisson's
equation
\begin{equation}  \label{Poisson}
\Delta V_e({\bf r})=\frac {e^2}{\varepsilon _0\varepsilon _i} \left( -n(%
{\bf r})+N_D({\bf r})-N_A({\bf r})\right),\quad i=1,2,
\end{equation}
where $\varepsilon_1$ and $\varepsilon_2$ are the dielectric constants of
the
semiconductor and oxide layers, respectively; $n({\bf r})$, $N_D({\bf r})$%
and $N_A({\bf r})$ are the concentrations of electrons, donors and acceptors
respectively. In our calculations we assume that the source electrode
is grounded whereas the potentials at the drain and gate electrodes
are equal $V_{ds}$ and $V_G$, respectively.

The study of the charge distribution in the cylindrical nanosize
MOSFETCGE structure in the state of the thermodynamical equilibrium
(see Ref.~\onlinecite{Pok}) has shown that the concentration of holes is
much lower than that of electrons so that electron transport is found to
provide the main contribution to the current flowing through the
MOSFET. For that reason, holes are neglected in the present transport
calculations.

\section{The Liouville equation}

In this section we consider ballistic transport of electrons. Neglecting
scattering processes and inter-valley transitions in the conduction band, the
one-electron density matrix can be written as
\begin{equation}
\rho({\bf r},{\bf r}^{\prime})=\sum_j \rho_j({\bf r},{\bf r}^{\prime}),
\end{equation}
where $\rho_j({\bf r},{\bf r}^{\prime})$ is the density matrix of electrons
residing in the $j$-th valley satisfying Liouville's equation
\begin{equation}  \label{liov1}
i\hbar \frac{\partial \rho_j }{\partial t}= \left[H_j,\rho_j \right].
\end{equation}
 In order to impose
reasonable boundary conditions for the density matrix in the electrodes, it
is convenient to describe the quantum transport along the $z$-axis in a
phase-space representation. In particular, we rewrite Eq.~(\ref{liov1})
in terms of $\zeta =(z+z^{\prime })/2$ and $\eta =z-z^{\prime }$
coordinates and express the density matrix $\rho _{j}$ as
\begin{equation}  \label{ser1}
\rho_j({\bf r},{\bf r})=\sum_{ms,m^{\prime}s^{\prime}} \frac 1{2\pi }%
\int_{-\infty} ^{+\infty }dke^{ik\eta }f_{jmsm^{\prime}s^{\prime}}(\zeta ,k)
\Psi_{jms}({\bf r}_\perp,z)\Psi^*_{jm^{\prime}s^{\prime}}({\bf r}%
^{\prime}_\perp,z^{\prime}),
\end{equation}
with a complete set of orthonormal functions $\Psi_{jms}({\bf r}_\perp,z)$.
According to the cylindrical symmetry of the system, these functions take
the following form:
\begin{equation}
\Psi_{jms}({\bf r}_\perp,z)=\frac 1{\sqrt{2\pi}}\psi_{jms}(r,z)e^{im\phi}.
\end{equation}
The functions $\psi_{jms}(r,z)$ are chosen to satisfy the equation
\begin{equation}  \label{radial}
-\frac{\hbar^2}{2\mjper}\left[\frac 1r\frac{\partial}{\partial r} \left(r%
\frac{\partial}{\partial r}\right)- \frac{m^2}{r^2}\right]\psi_{jms}(r,z)
+V(r,z)\psi_{jms}(r,z) ={\cal E}_{jms}(z)\psi_{jms}(r,z),
\end{equation}
which describes the radial motion of an electron. Here ${\cal E}_{jms}(z)$
are the eigenvalues of Eq.~(\ref{radial}) for a given value of the
$z$-coordinate which appears as a parameter. It will be
shown, that ${\cal E}_{jms}(z)$ plays the role of an effective potential in
the channel, and that $\Psi_{jms}({\bf r}_\perp,z)$ is the corresponding
wavefunction of the
transverse motion at fixed $z$.
Substituting the expansion (\ref{ser1}) into
Eq.~(\ref{liov1}), and using Eq.~(\ref{radial}), we arrive at an equation for
$f_{jmsm^{\prime}s^{\prime}}(\zeta ,k)$ :
\begin{eqnarray}
\label{liov3}
\frac{\partial f_{jmsm^{\prime}s^{\prime}}(\zeta,k)}{\partial t}&=& -\frac{%
\hbar k}{\mjpar}\frac{\partial}{\partial \zeta} f_{jmsm^{\prime}s^{%
\prime}}(\zeta,k)+\frac 1{\hbar}\int\limits_{-\infty}^{+\infty}
W_{jmsm^{\prime}s^{\prime}}(\zeta,k-k^{\prime})f_{jmsm^{\prime}s^{\prime}}(%
\zeta,k^{\prime})dk^{\prime}  \nonumber \\
&&-\sum_{s_1,s^{\prime}_1}\int\limits_{-\infty}^{+\infty} {\hat M}%
^{s_1s^{\prime}_1}_{jmsm^{\prime}s^{\prime}}(\zeta,k,k^{\prime})f_{jms_1m^{%
\prime}s^{\prime}_1}(\zeta,k^{\prime})dk^{\prime},
\end{eqnarray}
where
\begin{equation}
\label{w1}
W_{jmsm^{\prime}s^{\prime}}(\zeta,k-k^{\prime})=\frac{1}{2\pi i}%
\int\limits_{-\infty}^{+\infty} \left({\cal E}_{jms}(\zeta+\eta/2)-{\cal E}%
_{jm^{\prime}s^{\prime}}(\zeta-\eta/2)\right) e^{i(k^{\prime}-k)\eta}d\eta,
\end{equation}
\begin{equation}
{\hat M}^{s_1s^{\prime}_1}_{jmsm^{\prime}s^{\prime}}(\zeta,k,k^{\prime})=
\frac{1}{2\pi}\int\limits_{-\infty}^{+\infty} \left[\delta_{s^{\prime}s^{%
\prime}_1}{\hat \Gamma}_{mss_1}(\zeta+\eta/2,k^{\prime}) +\delta_{ss_1}{\hat %
\Gamma}^*_{ms^{\prime}s^{\prime}_1}(\zeta-\eta/2,k^{\prime}) \right]%
e^{i(k^{\prime}-k)\eta}d\eta,
\end{equation}
\begin{equation}
{\hat \Gamma}_{mss_1}(z,k^{\prime})= \frac{\hbar}{2\mjpar i}%
b_{jmss_1}(z) +\frac{\hbar}{2\mjpar}c_{jmss_1}(z) \left(-i\frac{%
\partial}{\partial \zeta}+2k^{\prime}\right),
\end{equation}
and
\begin{eqnarray}
b_{jmss_1}(z)&=& \int \psi^*_{jms}(r,z)\frac{\partial^2}{\partial z^2}%
\psi_{jms_1}(r,z)rdr, \\
c_{jmss_1}(z)&=& \int \psi^*_{jms}(r,z)\frac{\partial}{\partial z}%
\psi_{jms_1}(r,z)rdr.
\end{eqnarray}
Note that Eq.~(\ref{liov3}) is similar to the Liouville equation for the
Wigner distribution function, which is derived to model quantum
transport in tunneling diodes (see Ref.~\cite{Frensley}). The first
drift term in the right-hand side of Eq.~(\ref{liov3}) is derived from
the kinetic-energy operator of the longitudinal motion. It is exactly the
same as the corresponding term of the Boltzmann equation. The second
component plays the same role as the force term does in the Boltzmann
equation. The last term in the right-hand side of Eq.~(\ref{liov3}) contains
the operator ${\hat M}^{s_1s^{\prime}_1}_{jmsm^{\prime}s^{\prime}}(
\zeta,k,k^{\prime})$, which mixes the functions $f_{jmsm^{\prime}s^{\prime}}$
with different indexes $s,s^{\prime}$. It appears because $\psi_{jms}(r,z)$
are not eigenfunctions of the Hamiltonian (\ref{Hame}). The physical
meaning of the operator
${\hat M}^{s_1s^{\prime}_1}_{jmsm^{\prime}s^{\prime}}(\zeta,k,k^{\prime})$
will be discussed below.

In order to solve Eq.~(\ref{liov3}), we need to specify boundary conditions
for the functions $f_{jmsm^{\prime}s^{\prime}}(\zeta,k)$. For a weak
current, electrons incoming from both the source and the drain electrodes,
are assumed to be maintained in thermal equilibrium. Comparing
Eq.~(\ref{ser1}) with the corresponding expansion of the density matrix in the
equilibrium state, one obtains the following boundary conditions
\begin{eqnarray}
f_{jmsm^{\prime}s^{\prime}}(0,k)&=&\delta_{ss^{\prime}}\delta_{mm^{\prime}}2
\left[ \exp \left( E_{jsmk}\beta -E_{FS}\beta \right) +1\right] ^{-1},\qquad k>0,  \nonumber
\\
f_{jmsm^{\prime}s^{\prime}}(L,k)&=&\delta_{ss^{\prime}}\delta_{mm^{\prime}}2
\left[ \exp \left( E_{jsmk}\beta -E_{FD}\beta \right) +1\right] ^{-1},\qquad k<0,
\label{bound}
\end{eqnarray}
where the total energy is $E_{jsmk}=\hbar ^2k^2/2\mjpar+{\cal E}_{jsm}(0)$
for an electron
entering from the source electrode $(k>0)$ and $E_{jsmk}=\hbar
^2k^2/2\mjpar+{\cal E}_{jsm}(L)$ for an electron entering from the
drain electrode $(k<0)$. $E_{FS}$ and $E_{FD}$ are the Fermi energy levels in the source
and in the drain, respectively. Note, that Eq.~(\ref{bound}) meets the
requirement of imposing only one boundary condition on
the function $f_{jmsm^{\prime}s^{\prime}}(\zeta,k)$ at a fixed value of $k$
as Eq.~(\ref{liov3}) is a first order differential equation with respect
to $\zeta$.  Generally speaking, the solution of Eq.~(\ref{liov3}) with
the conditions (\ref{bound}) depends on the distance between the boundary
position
and the active device region. Let us estimate how far the boundary must be
from the active device region in order to avoid this dependence. It is easy
to show that the density matrix of the equilibrium state is a
decaying function of  $\eta =z-z^{\prime }$. The decay length is of the order
of the coherence length $\lambda_T=\sqrt{\frac{\hbar^2}{\mjpar k_BT}}
$ at high temperature and of the inverse Fermi wavenumber
$k_{F}^{-1}=\sqrt{{\hbar^2 }/{2\mjpar E_F}}$ at low temperature.
So, it is obvious, that
the distance between the boundary and the channel must exceed the
coherence length or the inverse Fermi wavenumber, i.~e. $L\gg \lambda_T$ or $%
L\gg k_{F}^{-1}$. For example, at $T=300$ K the coherence length $\lambda_T
\sim 3$ nm is much less than the source or drain lengths.

The functions $f_{jmsm^{\prime}s^{\prime}}(\zeta,k)$, which are introduced
in Eq.~(\ref{ser1}), are used in calculations of the current and the electron density. The
expression for the electron density follows directly from the density matrix
as $n({\bf r})=\rho({\bf r},{\bf r})$. In terms of the functions
$f_{jmsm^{\prime}s^{\prime}}(\zeta,k)$, the electron density can be written
as follows:
\begin{equation}  \label{nr}
n({\bf r})=\frac 1{2\pi}\sum_{jmsm^{\prime}s^{\prime}}
\int\limits_{-\infty}^{+\infty} f_{jmsm^{\prime}s^{\prime}}(z,k)dk
\Psi_{jms}({\bf r}_\perp,z)\Psi^*_{jm^{\prime}s^{\prime}}({\bf r}_\perp,z).
\end{equation}

\noindent
It is well-known~\cite{Landau}, that the current density can be expressed
in terms of the density matrix
\begin{equation}  \label{current}
{\bf j}({\bf r},t)=\sum_j \left.\frac{e\hbar}{2m_ji}\left(\frac{\partial}{%
\partial {\bf r}} -\frac{\partial}{\partial {\bf r}^{\prime}}\right)\rho_j(%
{\bf r},{\bf r}^{\prime},t) \right|_{{\bf r}={\bf r}^{\prime}}.
\end{equation}
The total current, which flows through the cross-section of the structure at
a point $z$, can be obtained by an integration over the transverse
coordinates. Substituting the expansion (\ref{ser1}) into Eq.~(\ref
{current}) and integrating over $r$ and $\varphi$, we find
\begin{equation}  \label{cur3}
J=e\sum_{j,m,s}\frac 1{2\pi}\int\limits_{-\infty}^{+\infty} dk \frac{\hbar k%
}{\mjpar} f_{jmsms}(z,k) -\frac{2e\hbar}{\mjpar}\sum_{%
%TCIMACRO{\QATOP{j,m,s,s^{\prime}}{s^{\prime}>s}}%
%BeginExpansion
{j,m,s,s^{\prime} \atop s^{\prime}>s}%
%EndExpansion
} c_{jmss^{\prime}}(z)\int\limits_{-\infty}^{+\infty} dk \ {\rm Im}
f_{jmsms^{\prime}}(z,k),
\end{equation}
where ${\rm Im} f$ is the imaginary part of $f$. The first term in the
right-hand side of Eq.~(\ref{cur3}) is similar to the expression for a
current of the classical theory~\cite{Landau}. The second term, which depends on the non-diagonal functions $%
f_{jmsms^{\prime}}$ only, takes into account the effects of intermixing
between different states of the transverse motion.

The last term in the right-hand side of Eq.~(\ref{liov3}) takes into
consideration the variation of the wavefunctions $\psi_{jms}(r,z)$ along the
$z$-axis. In the source and drain regions, the electrostatic potential is
essentially constant due to the high density of electrons. In these parts of the
structure, the wavefunctions of the transverse motion are very weakly
dependent on $z$, and consequently, the operator ${\hat M}^{s_1s^{
\prime}_1}_{jmsm^{\prime}s^{\prime}}$ has negligible effect. Inside the
channel, electrons are strongly localized at the Si/SiO$_2$ interface as the
positive gate voltage is applied. Earlier calculations, which we made for
the case of
equilibrium~\cite{Pok}, have shown that in the channel the dependence of
$\psi_{jms}(r,z)$ on $z$ is weak, too. Therefore, in the channel the effect
of the operator
${\hat M}^{s_1s^{\prime}_1}_{jmsm^{\prime}s^{\prime}}$ is negligible. In the
intermediate regions (the source--channel and the drain--channel), an
increase of the contribution of the third term in the right-hand side of
Eq.~(\ref{liov3}) is expected due to a sharp variation of $\psi_{jms}(r,z)$.
Since ${\hat M}^{s_1s^{\prime}_1}_{jmsm^{\prime}s^{\prime}}$ couples functions
$f_{jmsm^{\prime}s^{\prime}}(\zeta,k)$ with different quantum numbers
($jms$), it can be interpreted as
a collision operator, which describes transitions of electrons between
different quantum states of the transverse motion. Thus, the third term in the
right-hand side of Eq.~(\ref{liov3}) is significant only in the close vicinity
of the p-n junctions. Therefore, this term is assumed to give a
small contribution to the charge and current densities.
Under the above assumption, we have treated the last term in the right-hand
side of Eq.~(% \ref{liov3}) as a perturbation. Hereafter, we investigate the
steady state of the system in a zeroth order approximation with respect to
the operator  ${\hat M}^{s_1s^{\prime}_1}_{jmsm^{\prime}s^{\prime}}$.
Neglecting the latter, one finds that, due to the boundary conditions
(\ref{bound}), all non-diagonal functions
$f_{jmsm^{\prime}s^{\prime}}(\zeta,k)$ ($m\neq
m^{\prime}$ or $s\neq s^{\prime}$) need to be zero.

In the channel, the energy of the transverse motion can be approximately
written in the form \cite{Pok}
\begin{equation}
{\cal E}_{jms}(z)={\cal E}_{js}(z)+\frac{\hbar^2m^2}{2m^\perp_{js}
R^{*2}_{js}},
\end{equation}
where ${\cal E}_{js}(z)$ is the energy associated with the radial
size quantization and $%
\hbar^2m^2/2m^\perp_{js} R^{*2}_{js}$ is the energy of the angular motion
with averaged radius $R^{*}_{js}$.
Hence, in
Eq.~\ref{w1} for the diagonal functions $f_{jmsms}(\zeta,k)$
the difference ${\cal E}_{jms}(\zeta+\eta/2)-{\cal E}_{jms}(\zeta-\eta/2)$
can be substituted by ${\cal E}_{js}(\zeta+\eta/2)-{\cal E}_{js}(\zeta-\eta/2)$.
Furthermore, summation over $m$ in Eq.~(\ref{liov3}) gives
\begin{equation}
\label{liov6}
\frac{\hbar k}{\mjpar}\frac{\partial}{\partial \zeta} f_{js}(\zeta,k)-%
\frac 1{\hbar}\int\limits_{-\infty}^{+\infty}
W_{js}(\zeta,k-k^{\prime})f_{js}(\zeta,k^{\prime})dk^{\prime}=0
\end{equation}
with
\begin{equation}
f_{js}(\zeta,k)=\frac 1{2\pi}\sum_{m} f_{jmsms}(\zeta,k).
\end{equation}
In Eq.~(\ref{liov6}) the following notation is used
\begin{equation}
W_{js}(\zeta,k)=-\frac{1}{2\pi}\int\limits_{-\infty}^{+\infty} \left({\cal E}%
_{js}(\zeta+\eta/2)-{\cal E}_{js}(\zeta-\eta/2)\right) \sin(k\eta)d\eta.
\end{equation}
The effective potential ${\cal E}_{js}(z)$ can be interpreted as the bottom
of the subband $(j,s)$ in the channel.
The function $f_{js}(\zeta,k)$ is referred to as a partial Wigner
distribution function describing electrons which are travelling through the
channel in the inversion layer subband $(j,s)$.

\section{Electron scattering}

In this section we consider the electron scattering from phonons and impurities.
For this purpose we introduce a
Boltzmann-like single collision term~\cite{Landau}, which in the present
case has the following form
\begin{equation}
St \
f_{jsmk}=\sum_{j^{\prime}s^{\prime}m^{\prime}k^{\prime}}\left(P_{jsmk,j^{%
\prime}s^{\prime}m^{\prime}k^{\prime}}f_{j^{\prime}s^{\prime}m^{\prime}k^{%
\prime}}- P_{jsmk,j^{\prime}s^{\prime}m^{\prime}k^{\prime}}f_{jsmk}\right).
\label{1}
\end{equation}

As was noted above, we have neglected all transitions between quantum
states with different sets of quantum numbers $j$ and $s$. In the source
and drain contacts the distribution of electrons over the quantum states of
the angular motion corresponds to equilibrium. Consequently, due to the
cylindrical symmetry of the system, we may fairly assume that across the
whole structure the electron distribution is given by
\begin{equation}
f_{jsmk}(z)=f_{js}(z,k)w_{jsm},  \label{2}
\end{equation}
where
\begin{equation}
w_{jsm}=\sqrt{\frac{\hbar ^{2}\beta }{2\mjper R_{js}^{2}\pi }}\exp
\left( -\frac{\beta \hbar ^{2}m^{2}}{2\mjper R_{js}^{2}}\right)
\label{3}
\end{equation}
is the normalized Maxwellian distribution function with respect to the
angular momentum $m$.
The integration of the both sides of Eq.~(\ref{1}) over the angular
momentum gives the one-dimensional collision term
\begin{equation}
St\ f_{js}(z,k)=\sum_{k^{\prime }}\left( P_{js}(k,k^{\prime
})f_{js}(z,k^{\prime })-P_{js}(k^{\prime },k)f_{js}(z,k)\right) ,  \label{4}
\end{equation}
where
\begin{equation}
P_{js}(k,k^{\prime })=\sum_{mm^{\prime }}P_{jsmk,jsm^{\prime }k^{\prime
}}w_{jsm^{\prime }}.  \label{5}
\end{equation}
This collision term is directly incorporated into the
one-dimensional Liouville equation (\ref{liov6}) as
\begin{equation}
{\tilde W}_{js}(z,k,k^{\prime})=W_{js}(z,k-k^{\prime})+P_{js}(z,k,k^{%
\prime})- \delta_{k,k^{\prime}}\sum_{k^{\prime}}P_{js}(z,k^{\prime},k),
\end{equation}
where ${\tilde W}_{js}(z,k,k^{\prime})$ is the modified force term in
Eq.~(\ref{liov6}).

In this work we consider scattering by acceptor impurities and
acoustic phonons described by a deformation potential. The scattering rates
are evaluated according to Fermi's golden rule
\begin{equation}
P_{jsmk,jsm^{\prime}k^{\prime}}=\frac{2\pi}{\hbar} \left|\left\langle
jsm^{\prime}k^{\prime}\right|{\hat H}_{int} \left|jsmk\right\rangle\right|^2
\delta\left(E_{jsm^{\prime}k^{\prime}}-E_{jsmk}\right),
\end{equation}
where ${\hat H}_{int}$ is the Hamiltonian of the electron-phonon or the
electron-impurity interaction. Hereafter, we model the potential of an
ionized acceptor as
$U({\bf r})=4 \pi e^2 R_s^2/\varepsilon_1 \delta({\bf r})$,
where $R_s$ determines a cross-section for scattering by an impurity.
Consequently, the absolute value of the matrix element is
\begin{equation}
\left|\left\langle jsm^{\prime}k^{\prime}\right|U({\bf r-r}_i)
\left|jsmk\right\rangle\right| = 4 \pi e^2 R_s^2/ \varepsilon_1
\psi^2_{js}(r_i,z_i).
\end{equation}
Averaging this over a uniform distribution of acceptors results in
the following scattering rate
\begin{equation}  \label{pi}
P^i_{jsmk,jsm^{\prime}k^{\prime}}=C_i \int\limits_0^R \psi^4_{js}(r,z)
\delta\left(E_{jsm^{\prime}k^{\prime}}-E_{jsmkk}\right) r dr,
\end{equation}
where $C_i=N_a \left(4 \pi e^2 R_s^2/ \varepsilon_1 \right)^2/{\hbar}$
and $N_a$ is the acceptor concentration.

At room temperature the rate of the scattering by acoustic phonons has the
same form. Indeed, for $T=300$~K the thermal energy
$k_BT\gg \hbar \omega_{{\bf q}}$, therefore the acoustic deformation
potential scattering is approximately elastic, and the emission and
absorption rates are equal to each other.
For low energies we can  approximate the phonon number as
$N_{{\bf q}}\approx {k_BT}/{\hbar \omega_{{\bf q}}}\gg 1$ and the phonon
frequency $\omega_{{\bf  q}}=v_0q$, where $v_0$ is the sound velocity.
Assuming equipartition of energy in the acoustic modes, the scattering rate is
\begin{equation}
P^{ph}_{jsmk,jsm^{\prime}k^{\prime}}=\frac{2\pi}{V}C_{ph}\sum_{{\bf q}}
\left|\left\langle jsm^{\prime}k^{\prime}\right|e^{i{\bf q} \cdot {\bf r}}
\left|jsmk\right\rangle\right|^2
\delta\left(E_{jsm^{\prime}k^{\prime}}-E_{jsmk}\right),
\end{equation}
where the parameter $C_{ph}=4 \Sigma^2 k_B T/ 9\pi\rho v_0^2 \hbar$.
Integrating over ${\bf q}$ yields the scattering rate
$P^{ph}_{jsmk,jsm^{\prime}k^{\prime}}$ in the form~(\ref{pi}) with $C_{ph}$
instead of $C_{i}$. The full scattering rate
$P_{jsmk,jsm^{\prime}k^{\prime}}=P^{i}_{jsmk,jsm^{\prime}k^{\prime}}+
P^{ph}_{jsmk,jsm^{\prime}k^{\prime}}$ is then inserted into
Eq.~(\ref{5}) in order to obtain the one-dimensional scattering rate
\begin{equation}
P_{js}(z,k,k^{\prime})=\left(C_i+C_{ph}\right) a(z) F\left(\frac{\hbar^2
k^{\prime 2}}{2\mjpar}- \frac{\hbar^2 k^2}{2\mjpar}\right),
\end{equation}
where
\[
a(z) = \sqrt{\frac{2 m^\perp_j }{\hbar^2 \pi \beta}}
R_{js}(z)\int\limits_0^R \psi^4_{js}(r,z) rdr,
\quad  F(x)=e^{-x/2}{ K}_0(|x|/2),
\]
where $ K_0(x)$ is a McDonald function \cite{abram}.
 In calculations of the scattering by acoustic phonons the following values
 of parameters for Si are used: $\Sigma=$9.2~eV, $\rho=2.3283\cdot 10^3$~kg/m$^3$,
 $v_0=8.43\cdot 10^5$~cm/s \cite{Land}

\section{Numerical model}

The system under consideration consists of regions with high (the source
and drain) and  low (the channel) concentrations of electrons.
The corresponding electron distribution difference would produce a
considerable inaccuracy if we would have attempted to directly construct a
finite-difference analog of Eq.~(\ref{liov6}). It is worth
mentioning that, in the quasi-classical limit, i.~e. ${\cal E}%
_{js}(\zeta+\eta/2)-{\cal E}_{js}(\zeta-\eta/2)\approx \frac{\partial {\cal E%
}_{js}(\zeta)}{\partial \zeta} \eta$, Eq.~(\ref{liov6}) leads to the
Boltzmann equation with an effective potential which has the following
exact solution in the equilibrium state:
\begin{equation}
f^{eq}_{js}(\zeta,k)=\frac 1{\pi}\sum_{m}\left[ \exp \left( \frac{\hbar ^2k^2%
}{2\mjpar}\beta +{\cal E}_{j,s}(\zeta)\beta +\frac{\hbar^2m^2\beta%
}{2m^\perp_{js} R^{*2}_{js}(\zeta)}-E_{F}\beta \right) +1\right] ^{-1}
\label{equil}
\end{equation}
For numerical calculations it is useful to write down the partial Wigner
distribution function
as $f_{js}(\zeta,k)=f^{eq}_{js}(\zeta,k)+f^{d}_{js}(\zeta,k)$.
Inserting this into Eq.~(\ref{liov6}), one obtains the following
equation for $f^{d}_{js}(\zeta,k)$:
\begin{equation}  \label{liov61}
\frac{\hbar k}{\mjpar}\frac{\partial}{\partial \zeta}
f^{d}_{js}(\zeta,k)-\frac 1{\hbar}\int\limits_{-\infty}^{+\infty}
W_{js}(\zeta,k-k^{\prime})f^{d}_{js}(\zeta,k^{\prime})dk^{\prime}=B_{js}(%
\zeta,k),
\end{equation}
where
\begin{equation}  \label{z}
B_{js}(\zeta,k)=\frac{1}{2\pi}\int\limits_{-\infty}^{+\infty}dk^{\prime}\int%
\limits_{-\infty}^{+\infty}d\eta \left({\cal E}_{js}(\zeta+\frac{\eta}2)-{\cal E}%
_{js}(\zeta-\frac{\eta}2)- \frac{\partial {\cal E}_{js}(\zeta)}{\partial \zeta}
\eta\right) \sin\left[(k-k^{\prime})\eta\right]f^{eq}_{js}(\zeta,k^{\prime}).
\end{equation}
The unknown function $f^{d}_{js}(\zeta,k)$ takes
values of the same order throughout the whole system, and therefore
is suitable for
numerical computations. In the present work, we have used the
finite-difference model, which is described in Ref.~\onlinecite{Frensley}.
The position variable takes the set of discrete values $\zeta_i=\Delta\zeta i
$ for $\{i=0,\dots,N_\zeta\}$. The values of $k$ are also restricted to the
discrete set $k_p=(2p-N_k-1)\Delta k/2$ for $\{p=1,\dots,N_k\}$. On a
discrete mesh, the first derivative $\frac{\partial f_{js}}{ \partial
\zeta}(\zeta_i,k_p)$ is approximated by the left-hand difference for $k_p>0$
and the right-hand difference for $k_p<0$.  It was
shown In Ref.~\onlinecite{Frensley},
that such a choice of the finite-difference representation for the
derivatives leads to a stable discrete model. Projecting the equation
(\ref{liov61}) onto the finite-difference basis gives a matrix equation
${\bf L}\cdot{\bf f}={\bf b}$.
In the matrix ${\bf L}$, only the diagonal blocks and  one upper and
one lower co-diagonal blocks are nonzero:
\begin{equation}
{\bf L}=\left(
\begin{array}{ccccc}
A_{1} & -E & 0 & \ldots & 0 \\
-V & A_{2} & -E & \ldots & 0 \\
0 & -V & A_3 & \ldots & 0 \\
\vdots & \vdots & \vdots & \ddots & \vdots \\
0 & 0 & 0 & \ldots & A_{N_\zeta-1}
\end{array}
\right).
\end{equation}
Here, the $N_k\times N_k$ matrices $A_i$, $E$, and $V$ are
%\begin{equation}
\[
\left[A_i\right]_{pp^{\prime}}=\delta_{pp^{\prime}}-\frac{2\mjpar
\Delta\zeta}{\hbar^2(2p-N_k-1)\Delta k}W_{js}(\zeta_i,k_p-k_{p^{\prime}}),
%\end{equation}
\]
\begin{equation}
\left[E\right]_{pp^{\prime}}=\delta_{pp^{\prime}}\theta \left\{\frac{N_k+1}2%
-p\right\}, \quad \left[V\right]_{pp^{\prime}}=\delta_{pp^{\prime}}\theta
\left\{p-\frac{N_k+1}2\right\},
\end{equation}
and the vectors are
\begin{equation}
[f_i]_{p}=f_{js}(\zeta_i,k_p),\quad {\rm and}\quad
[b_i]_p=B_{js}(\zeta_i,k_p), \quad i=1,N_{\zeta-1} , \quad i=1,N_k.
\end{equation}

A recursive algorithm is used to solve the matrix equation ${\bf L}\cdot{\bf f}={\bf b}$.
Invoking downward elimination, we are dealing with
$B_i=\left(A_i-VB_{i-1}\right)^{-1}E$ and
$N_i=\left(A_i-VB_{i-1}\right)^{-1}\left(b_i+VN_{i-1}\right)$
$(i=1,\dots,N_\zeta)$ as relevant matrices and vectors.
Then, upward elimination eventually yields the solution
$f_i=B_if_{i+1}+N_i$ ($i=N_\zeta-1,\dots,1$).
If an index of a matrix or a vector is smaller than 1 or larger
than $N_\zeta-1$, the corresponding term is supposed to vanish.

In the channel, the difference between effective potentials
${\cal E}_{js}(\zeta)$ with different ($j,s$) is of the order of or larger
than the thermal energy $k_BT$.
Therefore, in the channel only a few lowest inversion subbands must be taken
into account. In the source and drain, however, many
quantum states $(j,s)$ of the radial motion are strongly populated by
electrons. Therefore, we should account for all of them in order to
calculate the charge distribution. Here, we can use the fact that, according
to our approximation, the current flows only through the lowest subbands
in the channel.
Hence, only for these subbands the partial Wigner distribution function of electrons is
non-equilibrium. In other subbands electrons are
maintained in the state of equilibrium, even when a bias is applied. So, in
Eq.~(\ref{nr}) for the electron density, we can substitute functions
$f_{jmsms}(z,k)$ of higher subbands by corresponding equilibrium functions.
Formally, adding and subtracting the equilibrium functions for the lowest
subbands in Eq.~(\ref{nr}), we arrive at the following equation for the
electron density
\begin{equation}  \label{nr1}
n({\bf r})=n_{eq}({\bf r})+\frac 1{2\pi}\sum_{js}
\int\limits_{-\infty}^{+\infty} dk \left[ f_{js}(z,k) \left
|\psi_{js}(r, z)\right|^2 -f_{js}^{eq}(z,k)\left|\psi_{js}^{eq}(r, z)
\right|^2 \right],
\end{equation}
where $n_{eq}({\bf r})$ and $\psi_{js}^{eq}(r,z)$
are the electron density and the
wavefunction of the radial motion in the state of equilibrium, respectively.
The summation on the right-hand side of Eq.~(\ref{nr1}) is performed only
over the lowest subbands. Since the electrostatic potential  does not
penetrate into the source and drain, we suppose that the
equilibrium electron density {\it in these regions} is well described by the
Thomas-Fermi approximation:
\begin{equation}  \label{nsd}
n_{eq}({\bf r})=N_C\frac 2{\sqrt{\pi }} F_{1/2}\left(
\beta\left(eV(r,z)+E_F-E_C\right)\right),
\end{equation}
where the Fermi integral is
\begin{equation}  \label{fint}
F_{1/2}(x)=\int_0^\infty \frac{\sqrt{t}dt}{\exp (t-x)+1}.
\end{equation}
Here $N_C$ is the effective density of states in the conduction band and $E_F$ is
the Fermi level of the system in the state of equilibrium.

\section{Numerical results}

During the device simulation three equations are solved
self-consistently: (i) the equation for the wavefunction of the radial motion
(\ref{radial}), (ii) the equation for the partial Wigner
distribution function (\ref{liov6})
and (iii) the Poisson equation (\ref{Poisson}). The methods of numerical
solution of Eqs.~(\ref{radial}) and (\ref{Poisson}) are the same as for the
equilibrium state~\cite{Pok}. The numerical model for Eq.~(\ref{liov6}) was
described in
the previous section. In the present calculations the four lowest subbands
$(j=1,2$ $s=1,2)$ are taken into account. The electron density in
the channel is obtained
 from Eq.~(\ref{nr}), whereas in the source and drain regions
it is determined from
Eq.~(\ref{nr1}). The calculations are performed for structures with a
channel of radius $R=50$~nm  and for various values of the length:
$L_{ch}=$40, 60, 70 and 80~nm. The width of the oxide layer is taken to be
4~nm. All calculations are carried
out with $N_{\zeta }=100$ and $N_{k}=100$.
The partial Wigner distribution
functions, which are obtained as a result of the self-consistent
procedure, are then used to calculate the current according to Eq.~(\ref
{current}).

We investigate two cases:
 ballistic transport and quantum transport.
The scattering of the electrons is taken into account.
The distribution of the electrostatic potential is represented in Fig.~2
for $V_{ds}=0.3$ V and $V_{G}=1$ V. This picture is typical for the MOSFET structure,
which is considered here. The cross-sections of the
electrostatic potentials for $r=0,30,40,45,48,50$~nm are shown in Fig.~3.
The main part of the applied gate voltage falls in the insulator (50~nm $<r<$
54~nm). Along the cylinder axis in the channel, the electrostatic potential
barrier for the electron increases up to about 0.4 eV.
Since the potential along the cylinder axis is always high, the current
mainly flows in a thin layer near the semiconductor-oxide interface.
This feature provides a  way of controlling $I_{ds}$ through the gate
voltage. Varying $V_{ds}$ and $V_{G}$ mainly changes the shape of this
narrow path, and, as a consequence, influences the form of the effective
potential ${\cal E}_{js}(z)$. As follows from Figs.~2~and~3, the radius
of the pillar can be taken shorter without causing barrier degradation.
At the p-n-junctions (source--channel and drain--channel) the
electrons meet barriers across the whole semiconductor.
These barriers are found to persist even for high values of the applied
source--drain voltage and prevent an electron flood from the side
of the strongly doped source. The pattern of the electrostatic
potentials differs mainly near the semiconductor-oxide interface, where the
inversion layer is formed. In Fig.~4 the effective potential for the lowest
inversion subband $(j=1,s=1)$ is plotted as a function of $z$
for different applied bias $V_{ds}=0,\dots ,0.5$~V, $V_{G}=1$~V, $L_{ch}=60$~nm.
It is seen that the effective potential reproduces the distribution of the
electrostatic potential near the semiconductor-oxide interface. In the case
of ballistic transport (dashed curve), the applied drain-source voltage
sharply drops near the drain-channel junction (Figs.~3~and~4). The
scattering of electrons (solid curve) smoothes out the applied voltage,
which is now varying linearly along the whole channel. Note, {\it
that the potential obtained by taking into account scattering is always
higher than that of the ballistic case}.
The explanation is clear from Fig.~5, where the linear electron
density is plotted for $V_{G}=1$~V and $L_{ch}=60$~nm. It is seen that, due
to scattering, the electron density in the channel (solid curve) rises and
smoothes out.
Hence, the applied gate voltage is screened more effectively, and as
a result, the potential exceeds that of the ballistic case.
It should be noted that at
equilibrium ($V_{ds}=0$) the linear density and the effective potential for both
cases (with and without scattering) are equal to each other. This result
follows from the principle of detailed balance.

The current-voltage characteristics (the current density $I=J/2\pi R$ vs.
the source--drain voltage $V_{ds}$)
are shown in Figs.~6~and~7 for the structures with
channel lengths $L_{ch}=40,\dots ,80$~nm. At a threshold voltage
$V_{ds}\approx 0.2$~V a kink in the $I$--$V$ characteristics of the device
is seen.  At subthreshold voltages $V_{ds}<0.2$~V the
derivative $dV_{ds}/dJ$ gives the resistance
of the structure. It is natural, that scattering enhances the resistance
of the structure (solid curve) compared to the ballistic transport (dashed
curve). Scattering is also found to smear the kink in the $I$--$V$
characteristic. At a voltage $V_{ds}>0.2$~V a saturation regime is reached.
In this part of $I-V$ characteristics, the current through the structure
increases more slowly than it does at a subthreshold voltage.
The slope of the $I-V$ curve in the saturation regime rises when
the length of the channel decreases. This effect is explained by a
reduction of the p-n junction barrier potential as the length of the
channel becomes shorter than the p-n junction width. In Fig.~7, one can see
that, when the transistor is switched off ($V_G<0.5$~V), the influence of
the scattering on the current is weak. This fact is due to a low
concentration of electrons, resulting in a low amplitude of the scattering
processes.

In Figs.~8a and 8b the contour plots of the partial Wigner distribution
function
($j=1,s=1$) are
given for both cases (a -- without and b -- with scattering). The lighter
regions in these plots indicate the higher density of electrons. Far from
the p-n-junction, where the effective potential varies almost linearly, the
partial Wigner distribution function can be interpreted as a distribution
of electrons in the phase space. When electrons travel in the inversion layer
without scattering, their velocity increases monotonously along the whole
channel. Therefore, in the phase-space representation the distribution of
ballistic electrons looks as {\it a narrow stream in the channel}
(Fig.~8a). As it is expected, {\it scattering washes out the electron jet
in the channel} (see Fig.~8b). It is worth mentioning that the electron
stream in the channel does not disappear. It means that in this
case the electron transport through the channel combines the features of
both diffusive and ballistic motion.

\section{Summary}

We have developed a model for the detailed investigation of quantum
transport in MOSFET devices. The model employs the Wigner distribution
function formalism allowing us to account for electron scattering by
impurities and phonons. Numerical simulation of a cylindrical nanosize
MOSFET structure was performed. $I-V$ characteristics for different values
of the channel length were obtained. It is shown that the slope of the
$I-V$ characteristic in the saturation regime rises as the channel length
increases. This is due to the decrease of the p-n junction barrier potential.

Finally, we have demonstrated that the inclusion of a collision term in
numerical simulation is important for low source--drain voltages. The
calculations have shown that the scattering leads to an increase of the
electron density in the channel and smoothes out the applied voltage along
the entire channel.
The analysis of the electron phase-space distribution in the channel
has shown that, in spite of scattering, electrons are able to flow through
the channel as a narrow stream although, to a certain extent, the
scattering is seen to wash out this jet. Accordingly, features of both
ballistic and diffusive transport are simultaneously encountered.

\acknowledgements
This work has been supported by the Interuniversitaire Attractiepolen
 --- Belgische Staat, Diensten van de Eerste Minister --
Wetenschappelijke, technische en culturele Aangelegenheden;
 PHANTOMS Research Network; F.W.O.-V. projects Nos. G.0287.95,
9.0193.97 and
W.O.G. WO.025.99N (Belgium).

\newpage
\bigskip

\begin{center}
{FIGURE CAPTIONS}
\end{center}

\bigskip

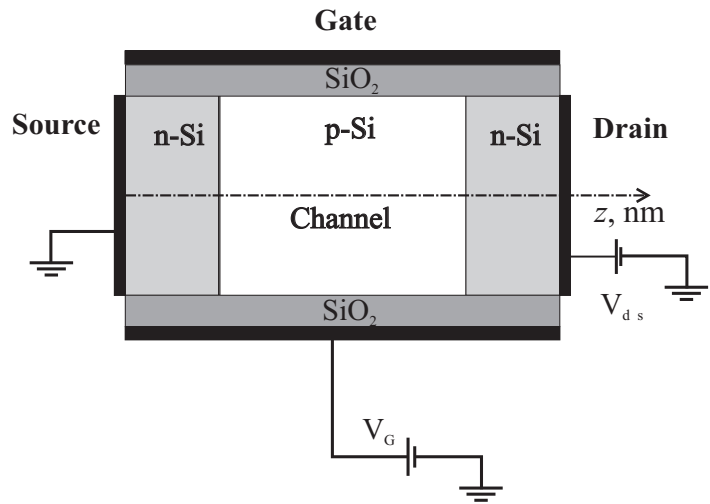
\noindent Fig. 1. Scheme of the cylindrical nanosize MOSFET.

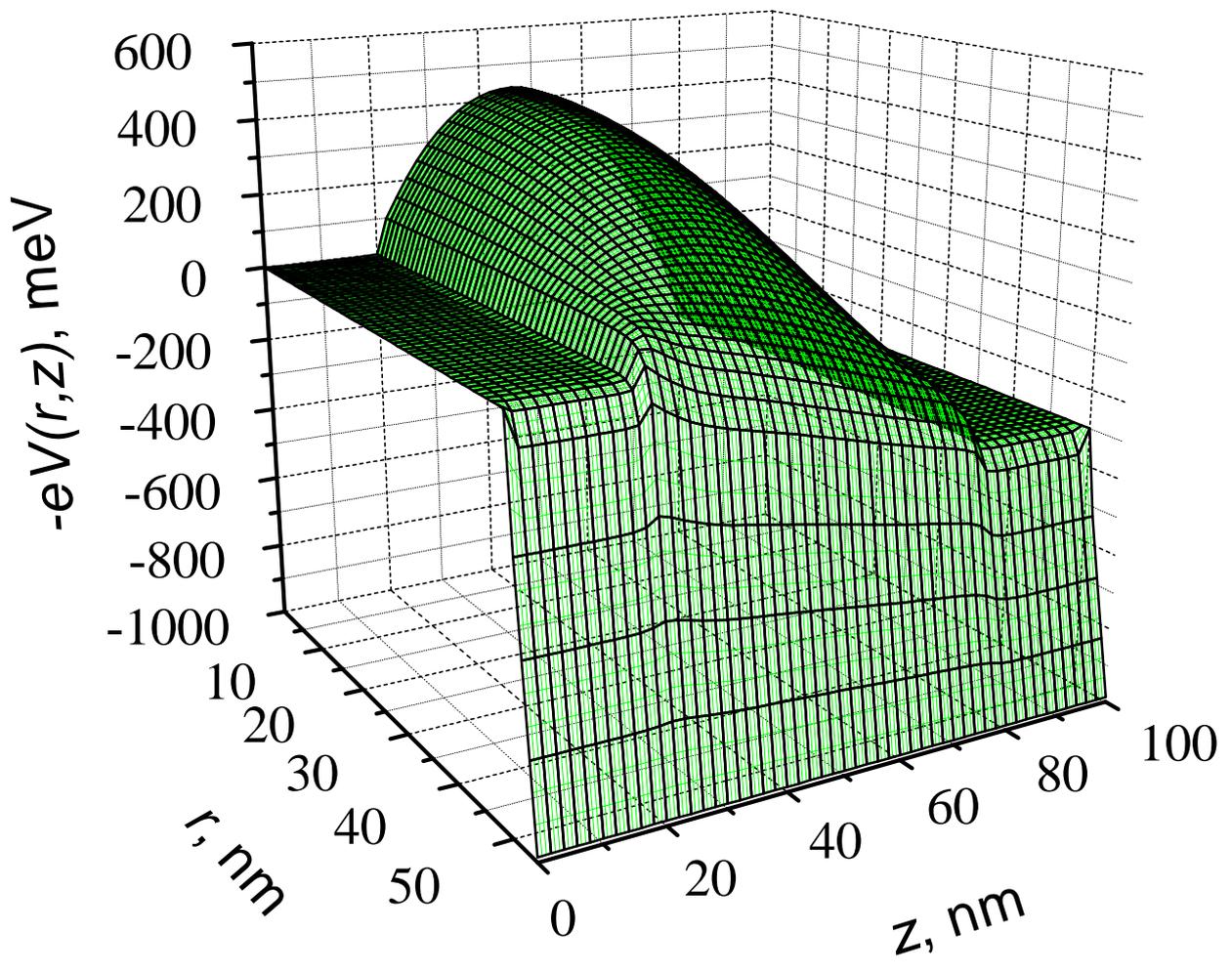
\noindent Fig. 2. Distribution of the electrostatic potential in the
MOSFET with $L_{ch}=60$~nm at $V_G=1$~V and $V_{ds}=0.3$~V.

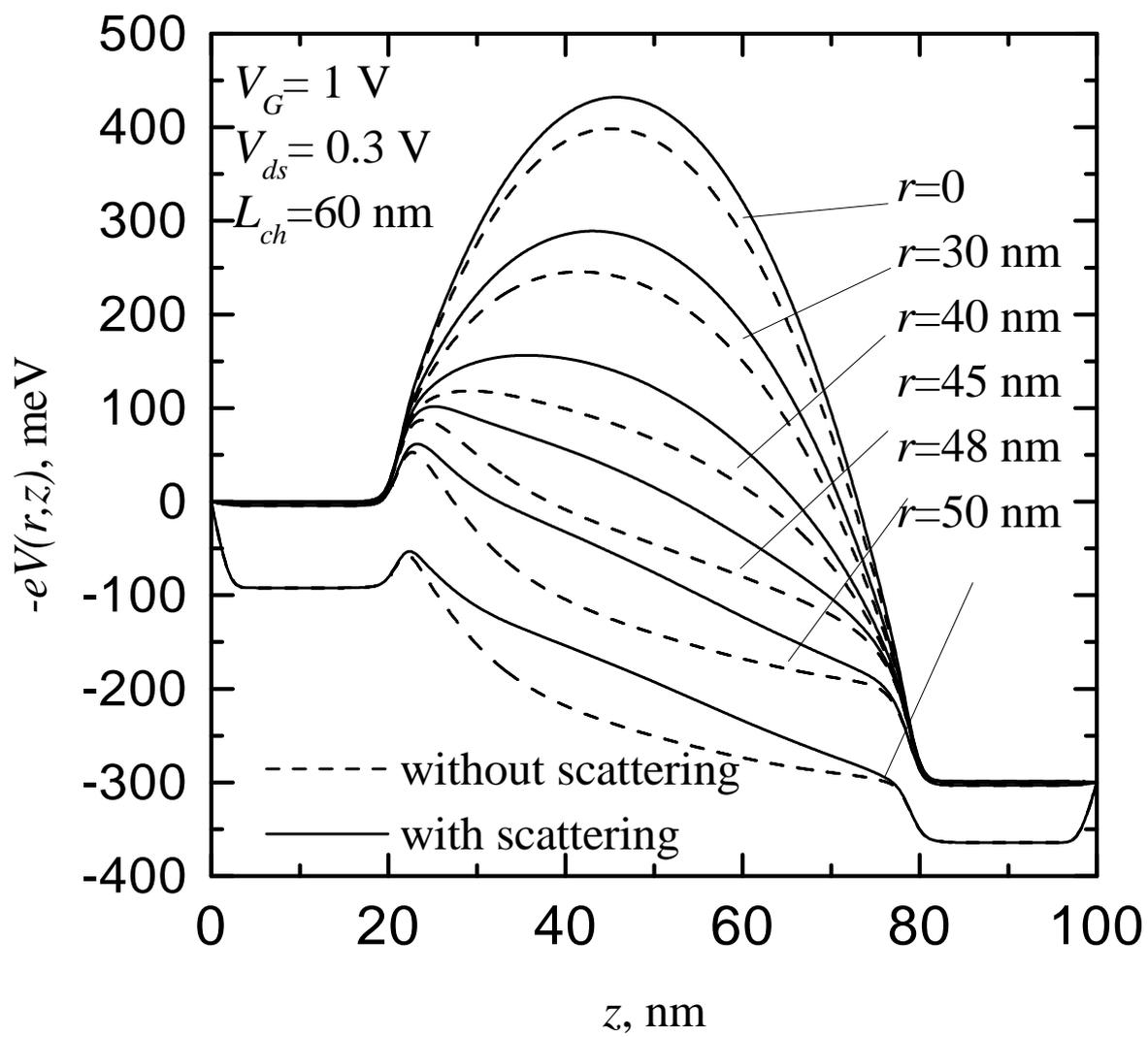
\noindent Fig. 3. Cross-sections of the electrostatic potentials
without scattering (dashed curves) and with scattering
from acceptor impurities and from an acoustic deformation potential
(solid curves).

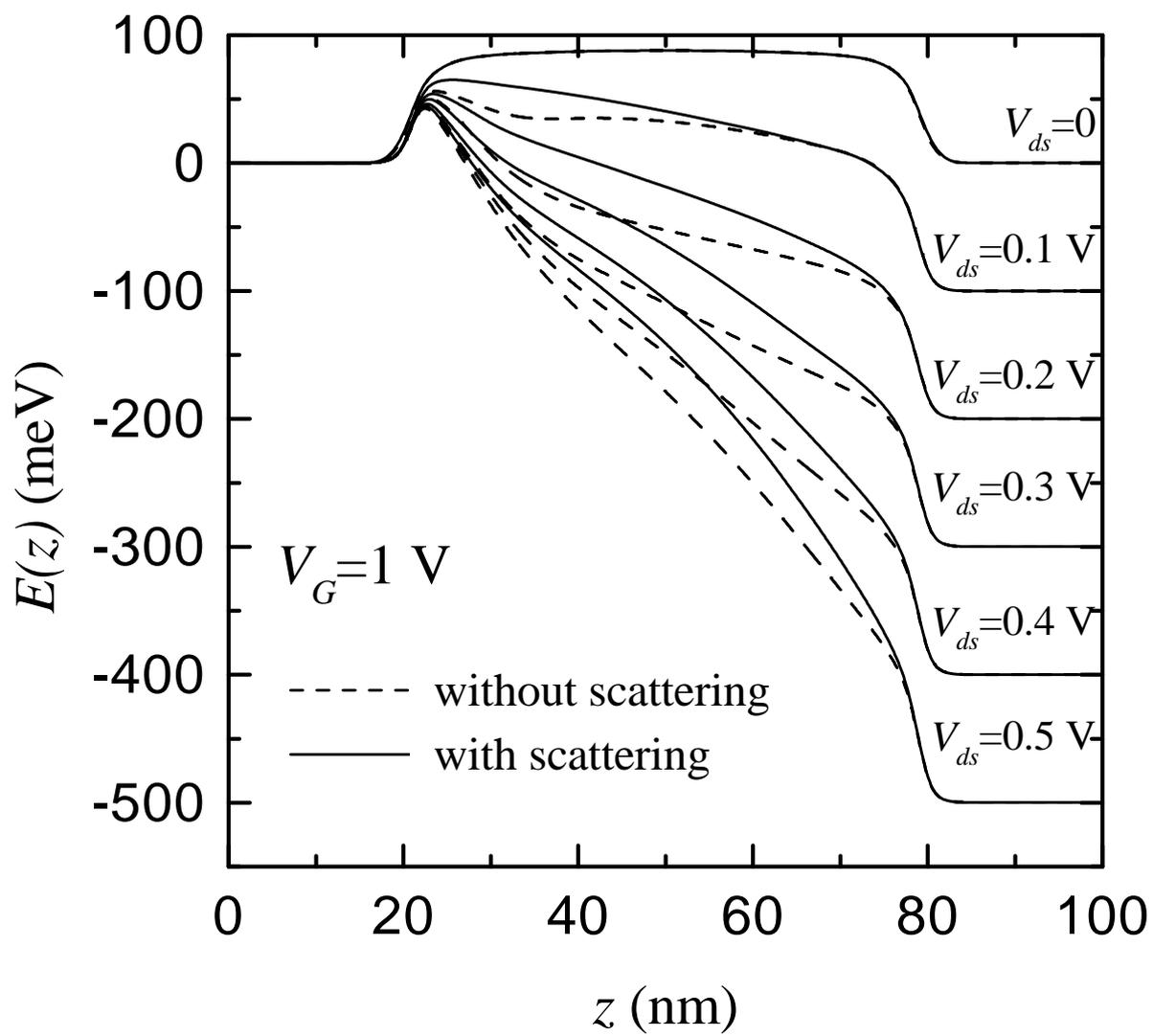
\noindent Fig. 4. Effective potential as a function of $z$	for
various $V_{ds}$, $L_{ch}=60$~nm.

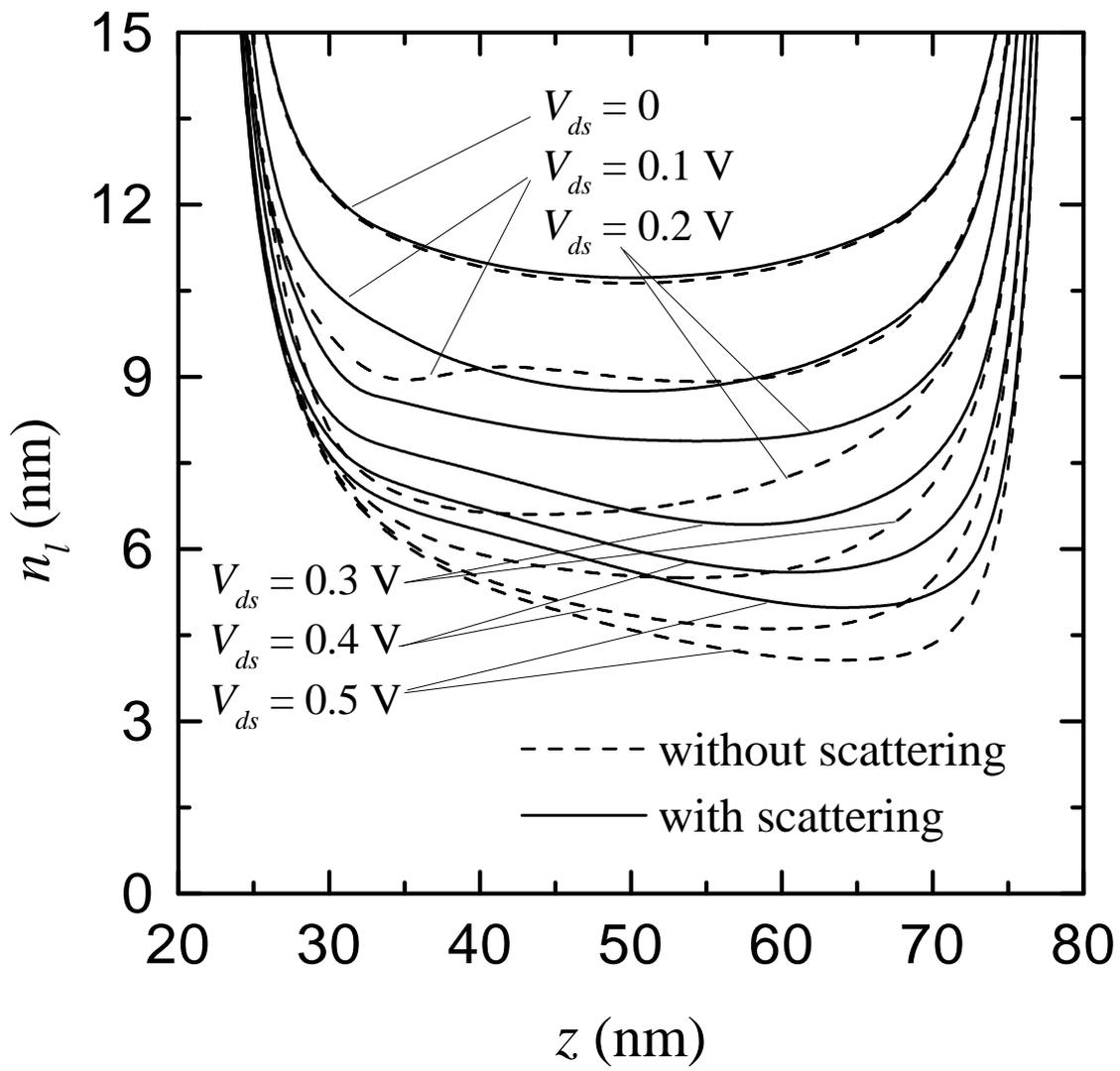
\noindent Fig. 5. Linear electron density in the channel as a function of $z$
for various $V_{ds}$, $L_{ch}=60$~nm.

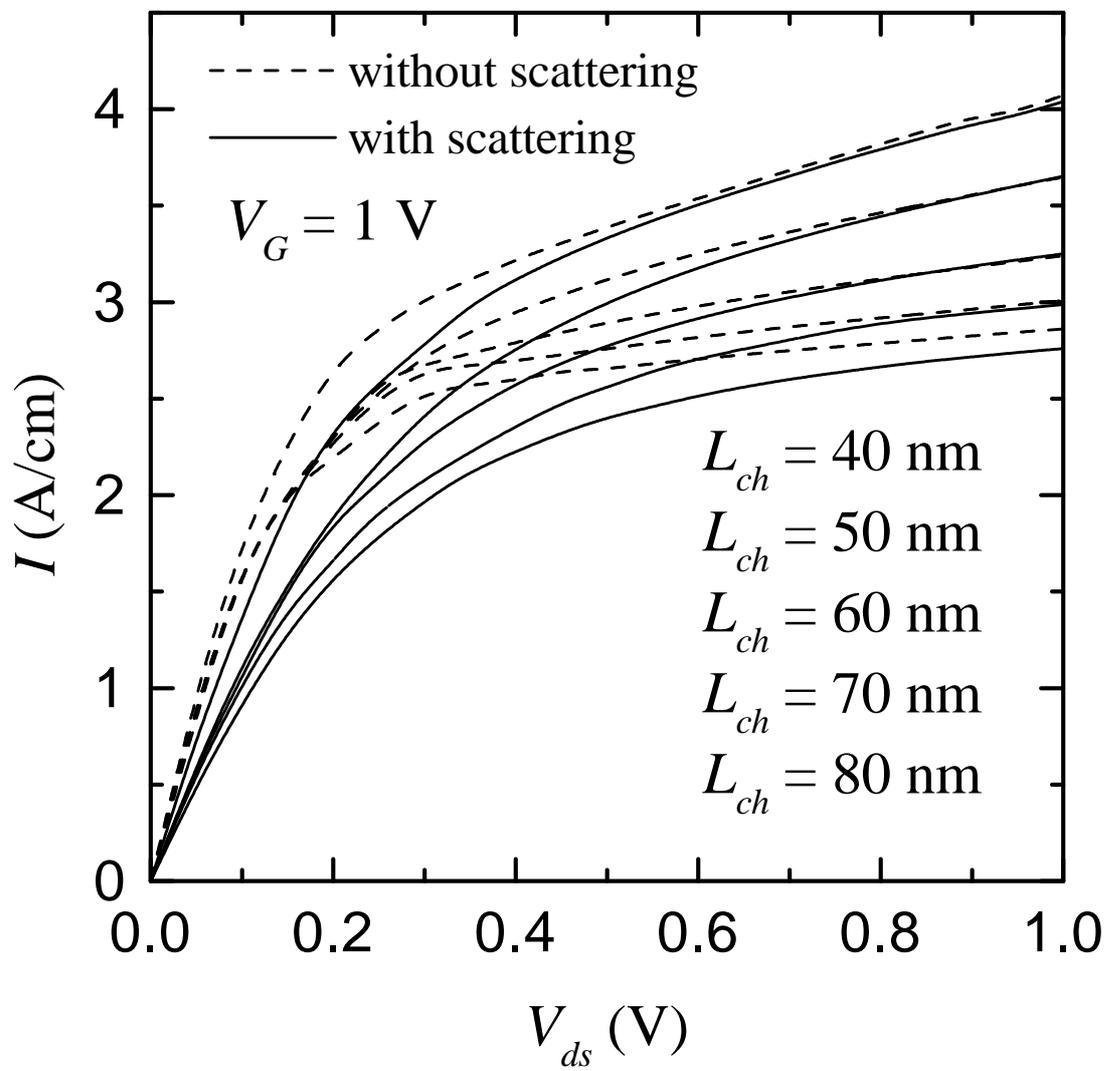
\noindent Fig. 6. Current-voltage characteristics at $V_G=1$~V for different
channel lengths.

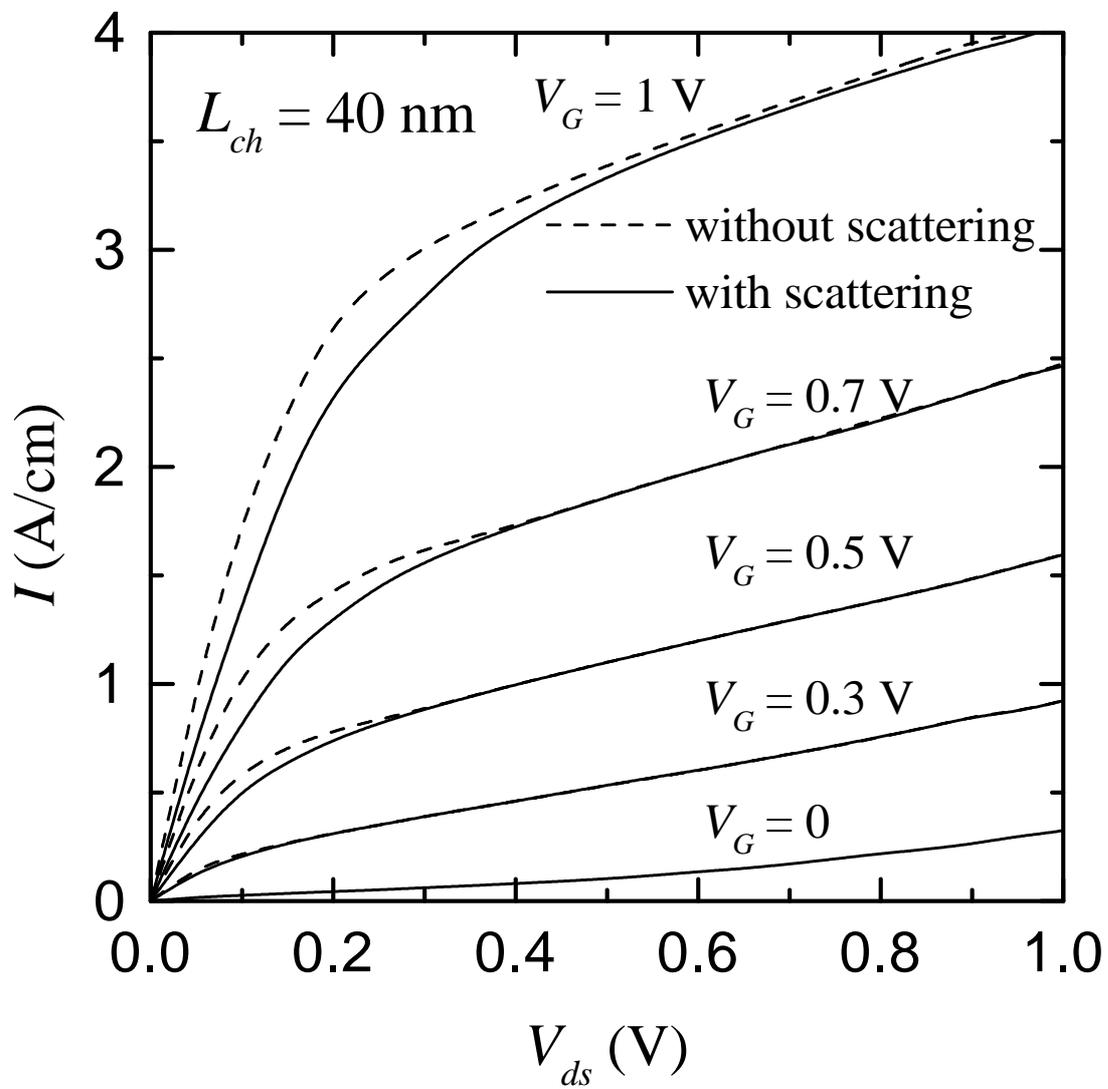
\noindent Fig. 7. Current-voltage characteristics for MOSFET with
$L_{ch}=40$~nm.

\noindent Fig. 8. Contour plots of the partial Wigner distribution function
$f_{js}(z,k)$ for
the lowest subband ($j=1,s=1$) at $V_G=1$~V, $V_{ds}=0.3$~V, $L_{ch}=60$~nm:
a -- without scattering, b -- with scattering.

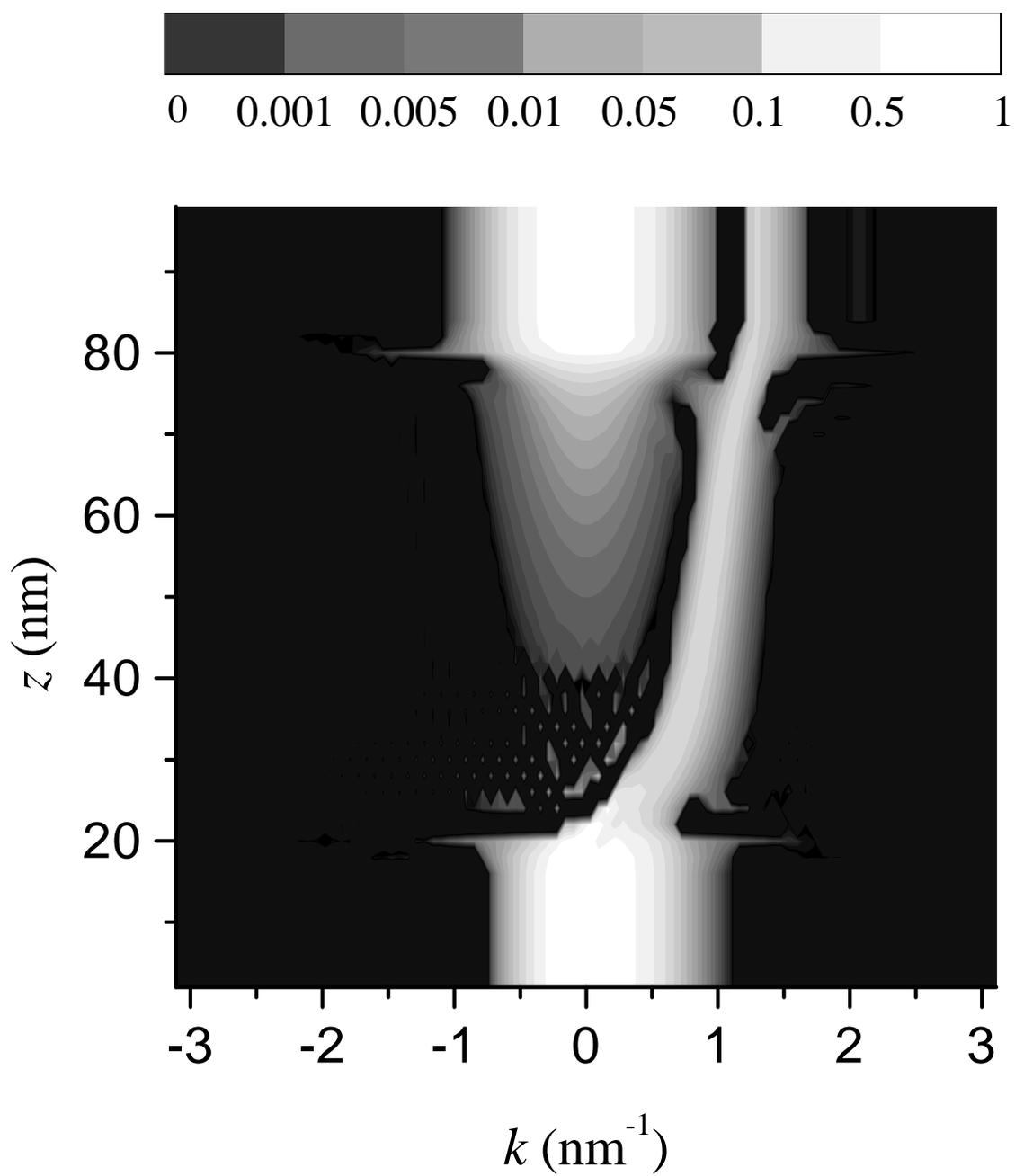
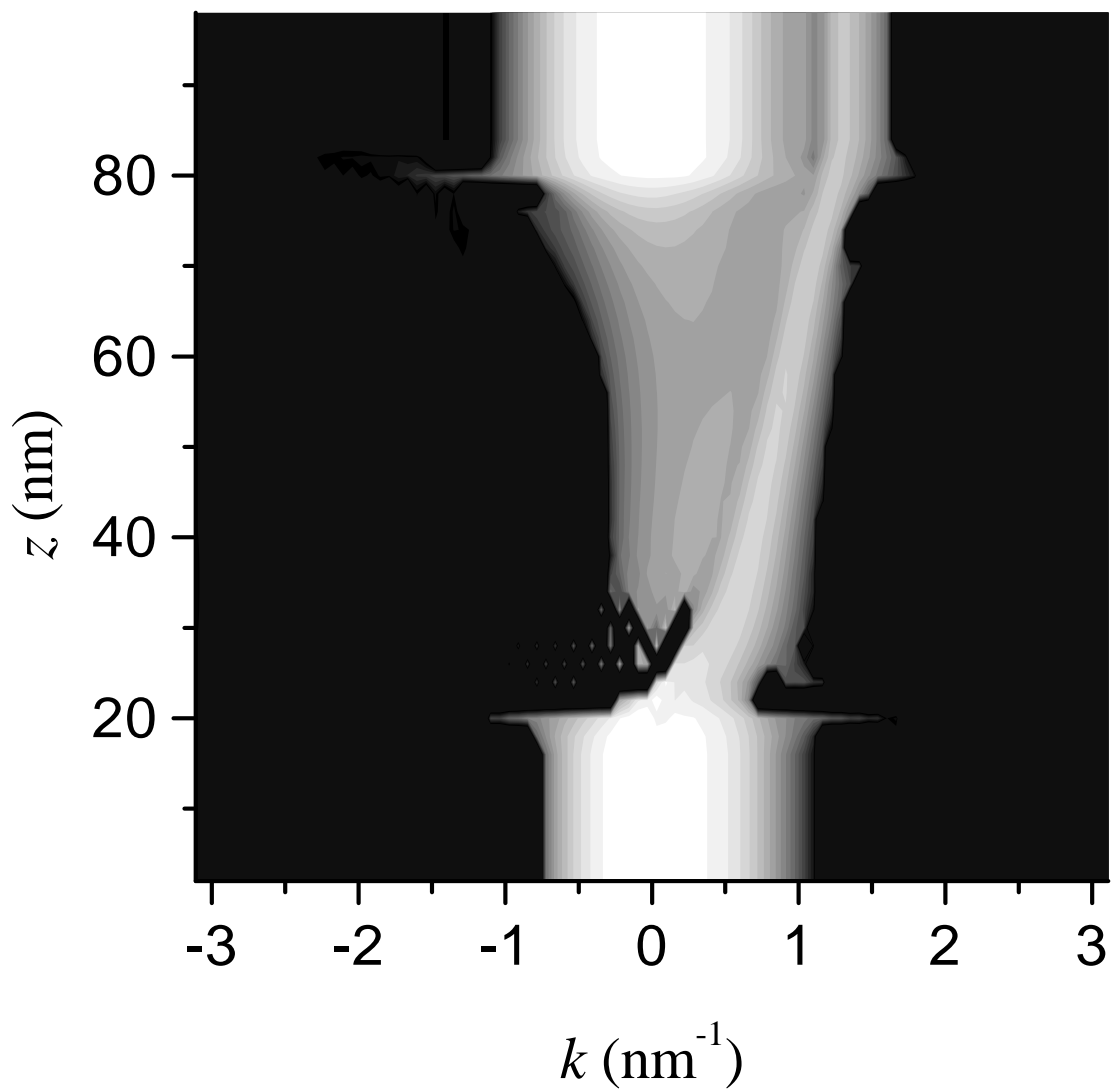
\end{document}